\documentclass[aip, rsi, reprint]{revtex4-1}

\usepackage{graphicx}
\graphicspath{{figs/}}
\usepackage{dcolumn}
\usepackage{bm}

\usepackage{amsmath}
\usepackage{amssymb}

\usepackage[utf8]{inputenc}
\usepackage[T1]{fontenc}
\usepackage{mathptmx}
\usepackage{etoolbox}
\usepackage{siunitx}

\makeatletter
\def\@email#1#2{%
 \endgroup
 \patchcmd{\titleblock@produce}
  {\frontmatter@RRAPformat}
  {\frontmatter@RRAPformat{\produce@RRAP{*#1\href{mailto:#2}{#2}}}\frontmatter@RRAPformat}
  {}{}
}%
\makeatother

\begin{document}

\preprint{RSI/E.Chebykin}

\title[Calibration of a Broadband Transmission Grating Spectrometer]{Absolute Calibration of a Broadband Transmission Grating Spectrometer from 5 to 400 nm with Uncertainty Estimates}

\author{E. Chebykin} 
\affiliation{Industrial Focus Group XUV Optics, MESA+ Institute for Nanotechnology, University of Twente, Drienerlolaan 5, 7522 NB, Enschede, The Netherlands}

\author{A.E.H. Gaballah}
\altaffiliation[Also at ]{Photometry Metrology Laboratory, Division of Photometry and Radiometry, National Institute of Standards (NIS), Tersa St, Al-Haram, 12211 Giza, Egypt}
\affiliation{Industrial Focus Group XUV Optics, MESA+ Institute for Nanotechnology, University of Twente, Drienerlolaan 5, 7522 NB, Enschede, The Netherlands}

\author{R. van der Horst}
\affiliation{ASML B.V., De Run 6501, 5504 DR, Veldhoven, The Netherlands}

\author{A. Giglia}
\author{N. Mahne} 
\altaffiliation[Also at ]{INAF - Osservatorio Astrofisico di Torino, Torino, Italy}
\affiliation{CNR-IOM – Istituto Officina dei Materiali, National Research Council of Italy, 34149 Trieste, Italy}

\author{M. van de Kerkhof} 
\affiliation{ASML B.V., De Run 6501, 5504 DR, Veldhoven, The Netherlands}

\author{M.D.~Ackermann}  
\altaffiliation[Also at ]{Advanced Research Center for Nanolithography (ARCNL), Science Park 106, 1098 XG Amsterdam, The Netherlands}
\affiliation{Industrial Focus Group XUV Optics, MESA+ Institute for Nanotechnology, University of Twente, Drienerlolaan 5, 7522 NB, Enschede, The Netherlands}

\author{M. Bayraktar} \email{m.bayraktar@utwente.nl}
\affiliation{Industrial Focus Group XUV Optics, MESA+ Institute for Nanotechnology, University of Twente, Drienerlolaan 5, 7522 NB, Enschede, The Netherlands}

\date{\today}

\begin{abstract}
We present the absolute calibration of a transmission grating spectrometer (TGS) with uncertainty estimates in the 5--400\,nm range using synchrotron radiation at the BEAR beamline at Elettra light source in Trieste, Italy. 
Firstly, the spectral responsivity of a Si photodiode in the spectrometer was absolutely calibrated with respect to the absolutely-calibrated beamline photodiode. 
Subsequently, this photodiode was used to calibrate the transmittance of Zr and Al thin-film filters, along with bulk MgF\textsubscript{2} and UV-grade fused silica windows at their operating positions in the spectrometer to minimize mounting-induced uncertainties. 
Next, the quantum efficiency of the spectrometer's CCD camera was calibrated over the 5--400\,nm range. 
Lastly, three gratings with line densities of \num{10000}, \num{3000}, and \num{1000}\,lines/mm were calibrated for spectroscopy in the ranges extending from soft X-ray to extreme-, vacuum- and deep-ultraviolet (DUV) wavelengths, respectively. 
The combination of various filters and gratings enables absolutely calibrated spectroscopy with high versatility across a broad wavelength range from soft X-rays to DUV. 
\end{abstract}


\maketitle 

\section{\label{sec:intro} Introduction}
Extreme ultraviolet (EUV) lithography is the leading technology for modern semiconductor manufacturing, using laser-produced tin (Sn) plasma as the light source\cite{banine2011, versolato2019}. 
The light originating from multiply charged Sn\textsuperscript{10+}--Sn\textsuperscript{15+} ions is mainly centered around 13.5\,nm\cite{churilov20061, churilov20062, churilov20063, osullivan2015, torretti2017, bouza2020, torretti2019, torretti2020}. 
The produced light is transmitted to a Si wafer through a series of optical components, the most critical of which are multilayer Mo/Si mirrors whose reflectance passband is precisely tuned to match the plasma emission around 13.5 nm\cite{louis2011, huang2017}. 
In addition to the dominant 13.5\,nm emission, the plasma also emits in a broad wavelength range from soft X-rays to vacuum-ultraviolet~(VUV) and deep-ultraviolet~(DUV)\cite{gambino2015, gambino2016, torretti2018, bouza2021}. 
In particular, radiation in the 30--90\,nm wavelength range is strongly absorbed by the hydrogen buffer gas in lithography scanners. 
Photoionization of hydrogen generates a secondary plasma that must be precisely controlled to prevent damage to the scanner components\cite{horst2016}. 
Furthermore, portions of the VUV and DUV radiation may be transported through the multilayer mirrors and reach the wafer surface. 
Given the broad wavelength range, the cumulative dose can be absorbed by the photoresist, causing parasitic exposure and lowering the contrast and quality of the image \cite{kerkhof20202, kerkhof20201}. 
Therefore, accurate characterization of the plasma emission spectrum in this broad wavelength range from soft X-rays to DUV is important for a better understanding of the  EUV emission mechanisms and developing potential improvement strategies for the lithography process.  

In recent years, spectrometers with free-standing transmission gratings have been used to measure plasma spectra over this broad wavelength range\cite{bouza2021}. 
Standard Si microfabrication techniques combined with nano-imprint or electron-beam lithography, enable patterning high line densities up to \num{10000}\, lines/mm, allowing sufficient spectral dispersion for soft X-ray and EUV wavelengths in a compact geometry\cite{Feng1997ATG, Fujikawa1998InSC, Sailaja1998ASX, Wilhein1999ASG, Weaver2001TheDO, Seely2001CalibrationOA, Hatsui2005DesignOA, Kornilov2010NanogratingbasedCV, ye2026}. 
Furthermore, combining gratings with different line densities on the same Si chip enables covering longer wavelengths using gratings with lower line densities, e.g.\,\num{1000}\,lines/mm, hence allowing to cover the soft x-rays to DUV range in a compact spectrometer design\cite{goh2015, bayraktar2016, bouza2021}. 
However, the efficiency of individual components of a transmission grating spectrometer (TGS), such as filters, gratings, and detectors, exhibits significant variations in the broad wavelength range covered. 
Previous intensity calibration efforts revealed that there are large variations, up to a factor of two in certain wavelength regions, among the reported efficiency curves of the CCD camera used in the spectrometer\cite{bouza2021}. 
Furthermore, rigorous uncertainty quantification is not available, which limits both the accuracy of the absolute calibration capability and the reliability of the interpretation of the measured spectra. 

Here, we present the absolute calibration of a TGS with uncertainty estimates in the 5-400\,nm range using synchrotron radiation at the BEAR beamline at the Elettra light source in Trieste, Italy\cite{Nannarone2004, Pasquali2004}. 
The transmission geometry allows a greater tolerance to variations in the angle of incidence compared to reflective spectrometers, thereby enabling a robust component-by-component calibration.
Specifically, using the absolute calibration of the beamline photodiode that can be traced to a primary calibration standard, we have initially calibrated the photodiode integrated in the TGS. 
The calibrated TGS photodiode was employed for the  calibration of filters. 

\section{\label{sec:elettra} Experimental procedures}

The BEAR beamline is equipped with a monochromator, based on a normal incidence channel for the UV range 2.8--40\,eV and a grazing incidence channel for the EUV-soft X-ray range 40--1600\,eV. Such beamline configuration delivers a maximum photon flux of $2\times10^{11}$\,photons/s in the EUV range. The beamline exhibits predominantly linear polarization in the \textit{s-} (horizontal) plane, with a polarization degree greater than 90\% \cite{Nannarone2004}. 
The component to be calibrated can be rotated by $90^{\circ}$ changing the incidence configuration from \textit{s-} (horizontal) to \textit{p-} (vertical) polarization plane \cite{Pasquali2004}. 
A set of filters is employed to suppress higher diffraction orders of the monochromators, ensuring a well-defined photon flux at the selected energy. 
For this calibration campaign, the beamline has been operated at its standard flux level, with the exit slits set to 900$\times$450\,\textmu m\textsuperscript{2}, in width and height respectively, to define the beam profile in the 2.8--18\,eV range.
For higher photon energies in the 15--250\,eV range, the exit slit height was reduced to 900$\times$100\,\textmu m\textsuperscript{2} to reduce the incident photon flux and prevent saturation of the CCD camera. 
The beam spot measures 4.13$\times$3.16\,mm\textsuperscript{2} and 3.19$\times$2.0\,mm\textsuperscript{2} for standard and low flux configurations, respectively. 
The beam intensity is monitored by reading the current of the last mirror of the beamline to correct for measurement-to-measurement intensity variations.  

\begin{figure}[tbh]
\includegraphics[width=\columnwidth]{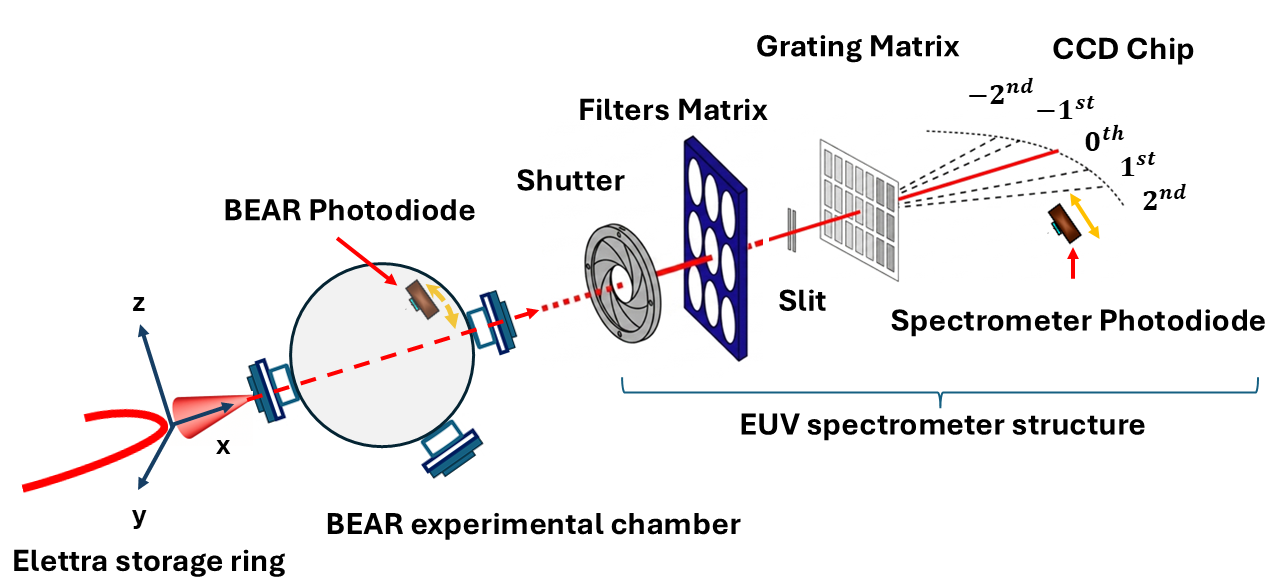}%
\caption{\label{fig:scheme} Schematic of spectrometer calibration setup at the BEAR beamline.}
\end{figure}

The beamline is nominally aligned with an experimental chamber that is equipped with an absolutely calibrated photodiode  (AXUV100G, DigiKey) that can be retracted as schematically shown in Fig.\,\ref{fig:scheme}. 
The spectrometer was mounted after the experimental chamber. 
The spectrometer includes a shutter, a filter matrix, a pair of slits, a grating chip produced in-house in the MESA+ cleanroom\cite{goh2015, bayraktar2016}, a photodiode (AXUV20HS1, DigiKey) and a CCD camera (PIXIS-XO:2KB, Princeton Instruments) as shown in Fig.\,\ref{fig:scheme}. 
A common trigger signal was used to synchronize the beamline hardware, the shutter, and the CCD camera. 
The spectrometer components were mounted on translation stages to enable precise alignment and movement in and out of the beam path. 

\section{\label{sec:PD} Photodiode calibration}
To assess the transmittance of the filters, we calibrated the spectrometer photodiode against the BEAR beamline's reference photodiode. 
The spectral responsivity of the photodiode is determined from the signals recorded by both photodiodes using the relation: 
\begin{equation}
    R_{spec}(\lambda)=R_{ref}(\lambda)\frac{I_{spec}(\lambda)}{I_{ref}(\lambda)}
\end{equation}
where $R(\lambda)$ and $I(\lambda)$ represent, respectively, the spectral responsivity and diode currents for the reference (\textit{ref}) and spectrometer (\textit{spec}) photodiodes. 
It was made sure that beam clipping was prevented in both photodiode measurements, dark current correction was applied, and time-dependent photon flux variations were corrected by normalizing photodiode signals with respective monitor currents (not shown). 

\begin{figure}
\includegraphics[width=\columnwidth]{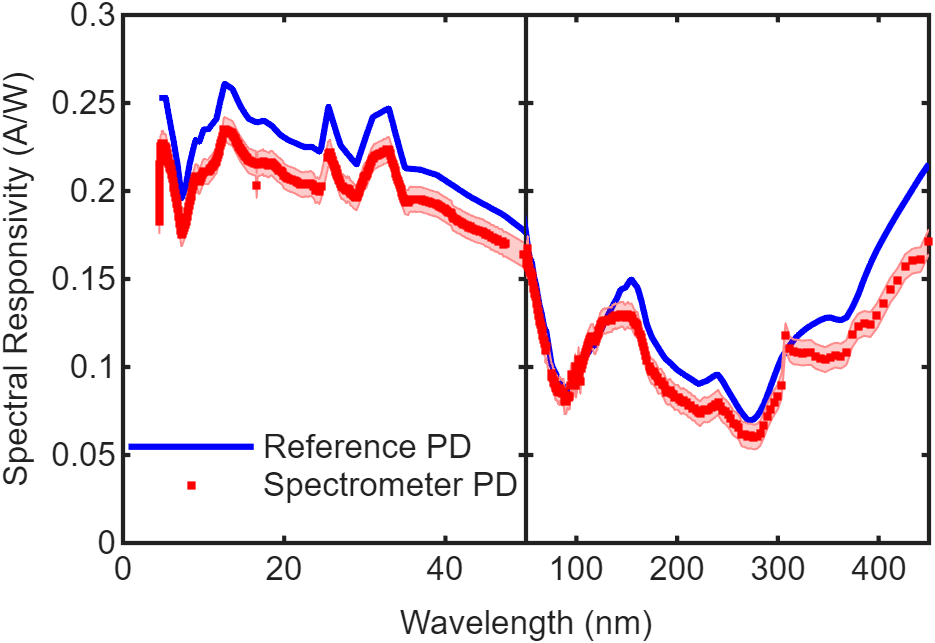}%
\caption{\label{fig:responsivity} Spectral responsivity of reference and spectrometer photodiodes (PDs). }
\end{figure}

Fig.~\ref{fig:responsivity} shows the spectral responsivity of the reference and spectrometer photodiodes, along with the responsivity from the manufacturer\,\cite{AXUV_datasheet}. 
Here and in the rest of the paper, the error bands are calculated as 3$\%$ of the maximum photodiode signal in each filter range, based on experiments performed at this beamline in Ref.~\onlinecite{thesisTryus}. 
There is an overall matching of the reference and spectrometer photodiode responsivities, especially in the 50--150\,nm range. Yet, some differences in the rest of the wavelengths are visible, possibly due to the fabrication process of the two photodiodes. 

\section{\label{sec:filters} Filter transmittance}
\begin{figure}[htb]
\includegraphics[width=\columnwidth]{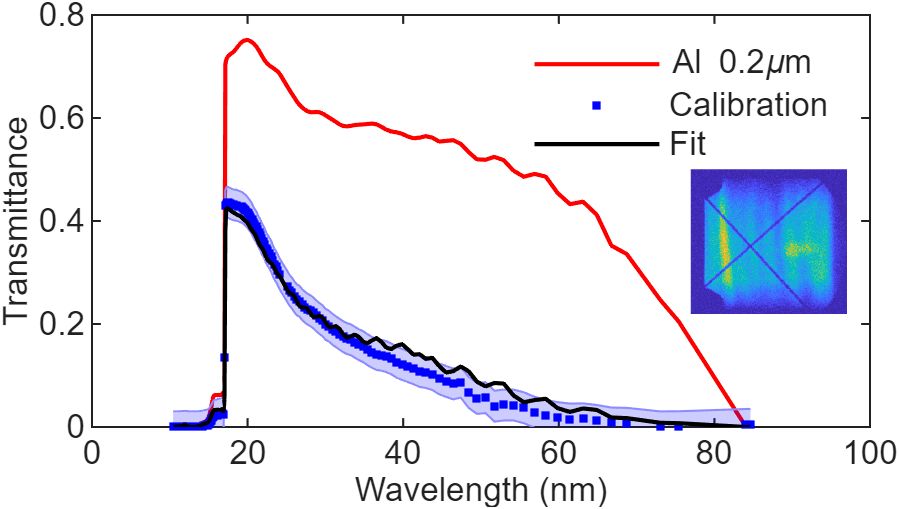}%
\caption{\label{fig:Zr200} EUV filter transmittance, Zr - 0.2\,\textmu m (top) , Al - 0.2\,\textmu m (bottom). Inset: beam profile, image size is 4.3\,mm$\times$3.9\,mm.}%
\end{figure}

\begin{figure}[htb]
\includegraphics[width=\columnwidth]{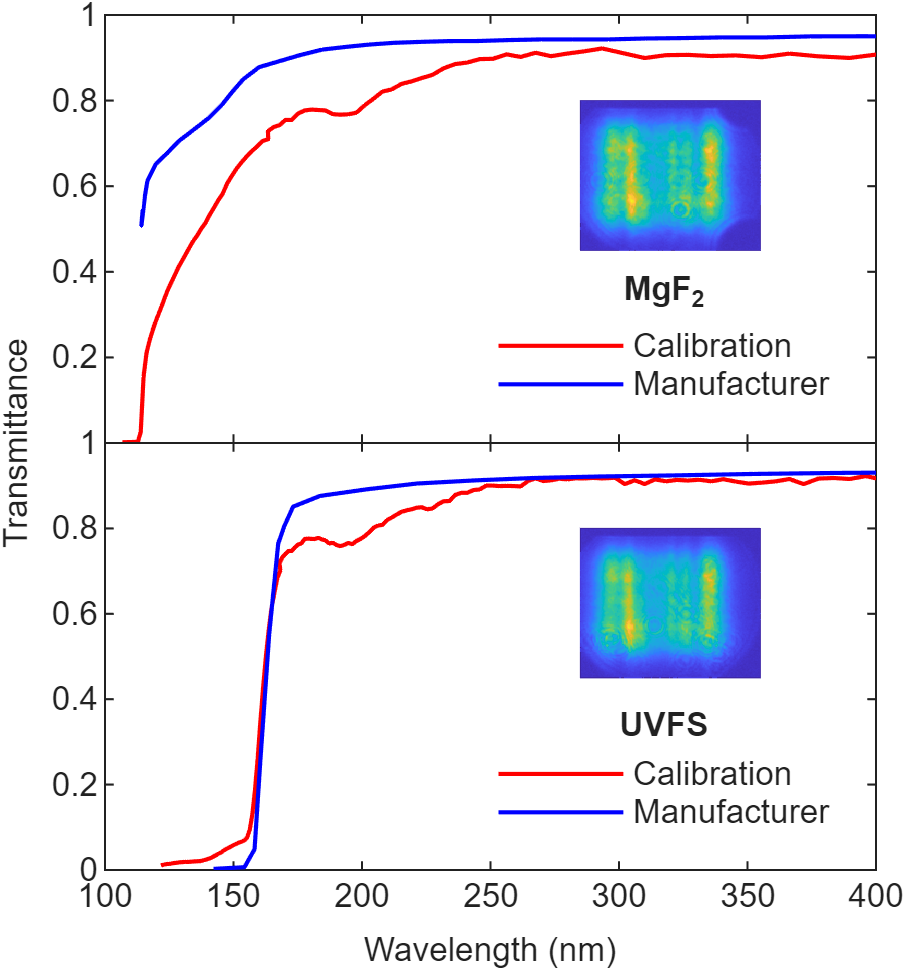}%
\caption{\label{fig:UVFS} VUV filter transmittance, MgF\textsubscript{2} (top), UVFS (bottom). Inset: beam profile, image size is 4\,mm$\times$3.6\,mm.}%
\end{figure}

Several filters are used in the spectrometer to eliminate the overlap of higher diffraction orders from strong short-wavelength emission with longer wavelengths. 
In certain measurements, filters are also employed to limit the measured spectral range to enhance the level of signal-to-noise ratio. 
In this measurement campaign, the following filters have been calibrated; EUV foil filters of aluminum (Al) and zirconium (Zr) with 0.2\,\textmu m nominal thickness (Lebow Inc.) and VUV filter windows of MgF\textsubscript{2} (Korth Kristalle) and UV-grade fused silica (UVFS, Thorlabs) with 2\,mm thickness. 
The transmittance of the filters was determined using the spectrometer's photodiode by taking the ratio of a measurement with the filter on the beam path to a measurement after removing the filter from the beam path.  
The measured transmittance are shown in Figs.\,\ref{fig:Zr200}--\ref{fig:UVFS} including the insets that show the beam profiles. 

\begin{figure}[htb]
\includegraphics[width=\columnwidth]{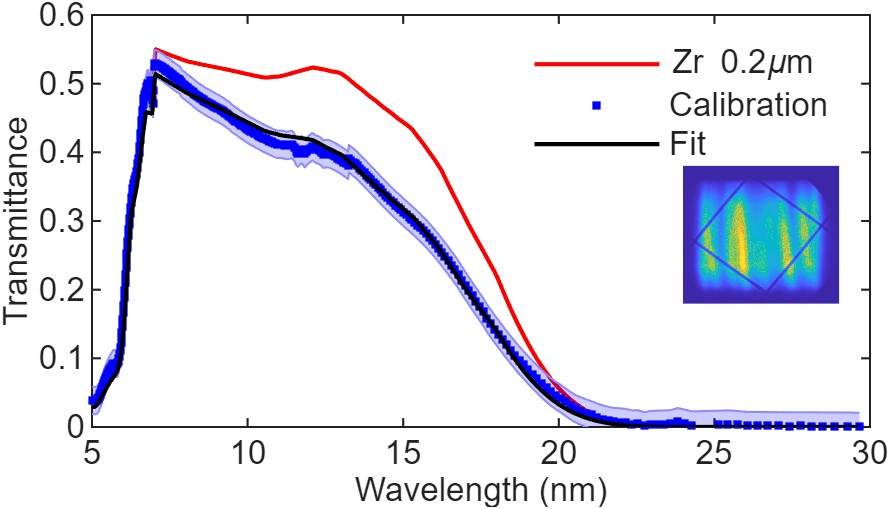}%
\end{figure}

Fig.\,\ref{fig:Zr200} compares the measured foil transmittance with the expected transmission assuming nominal foil thickness and using refractive index data reported in the literature\cite{palik1985, henke1993, Burcklen2024}. 
The error bands were assumed to be 3$\%$ of the maximum transmittance. 
It is observed that the two transmittance curves differ significantly from each other. This difference is likely due to the initial thickness of the foils being different from the nominal thickness and also due to the oxidation of the foils. 
To estimate the thickness of the oxide layer, the oxidation was assumed to occur uniformly on both sides of the foil, resulting in an oxide-metal-oxide stack. 
To model the transmission of the stack, we used the transfer matrix method (TMM)\cite{born1999, mackay2020} together with optical constants for metals (Al and Zr) and their corresponding oxides (Al\textsubscript{2}O\textsubscript{3} and ZrO\textsubscript{2})\cite{palik1985, henke1993, Burcklen2024}. 
We used the TMM in a non-linear least-squares fit algorithm to determine both the initial thickness of the foil and the oxide layer thickness. 
As a constraint, we considered that the total amount of metal in the foil is conserved after oxidation and equal to the initial thickness ($t_{Ini}$). Consequently, the total amount of metal in the oxidized foil can be expressed as a sum of the thickness of the non-oxidized metal part ($t_{Me}$) and the fraction of metal within the oxide layer ($t_{Ox}$). 
Mathematically, it can be written as:
\begin{equation}
    t_{Me}+\frac{\alpha}{\beta} \frac{M(Me)}{M(Ox)}t_{Ox}=t_{Ini}
\end{equation}
where $M$ -- molecular weight of the metal/oxide molecule, and $\alpha$ and $\beta$ are the stoichiometric coefficients in the oxidation equation for metal and oxide, respectively. 
As an example, for Al the oxidation equation is 4\,Al\,+\,3\,O\textsubscript{2}\,=\,2\,Al\textsubscript{2}O\textsubscript{3}, so $\alpha=4$ and $\beta=2$. 
Using this approach, it is possible to fit the TMM predictions to measurements using two free fit parameters, namely, the initial thickness and the oxide layer thickness. 

The fit results shown in Fig.~\ref{fig:Zr200} agree well with the measurements.
Table~\ref{tab:oxide_thickness} summarizes the thicknesses of the fitted layers and the quality of fit using reduced $\chi^2$. 
The values of $\bar{\chi}^2$ are significantly lower than 1, due to the large number of data points, and also indicating that the error-bands used are overestimating the actual uncertainty. 
However, we keep the error-bands at 3\% as an upper bound for uncertainty. 

The initial fitted thicknesses are around 180\,nm for both foils, approximately 10\% lower than the nominal 200\,nm thickness, either due to actual manufacturing tolerance or reduced density of the foils. 
The estimated total oxide thickness agrees well with published studies. 
Ref.\,\onlinecite{joseph20131} reports about 15\,nm thick oxide layer on an Al foil, which is in agreement with our results of  18.5\,nm. Ref.\,\onlinecite{powell1990} demonstrates oxidation of Zr foils by using depth profiling and reports a total of 20\,nm thick oxide on a 100\,nm thick foil. 
Our total thickness of the ZrO\textsubscript{2} layer is in the 30\,nm range, having a moderate agreement with this study. 

\begin{table}
\centering
\caption{Estimation of layer thicknesses}
\label{tab:oxide_thickness}
\renewcommand{\arraystretch}{1.4} 
\begin{tabular}{lccccc}
\hline
Filter & $t_{Ini}\,(nm)$ & $t_{Me}$\,(nm) & $t_{Ox}$\,(nm) & $t_{Tot}$\,(nm) & $\bar{\chi}^2$ \\
\hline
Al & $180.5 \pm 2.5$ & $170.5 \pm 2.5$ & $18.5 \pm 3.5$ & $189 \pm 4$ & 0.18 \\
Zr & $184.0 \pm 1.0$ & $161.5 \pm 1.3$ & $30.5 \pm 1.5$ & $192 \pm 2$ & 0.015 \\
\hline
\end{tabular}
\renewcommand{\arraystretch}{1} 
\end{table}

The calibrated transmittance for the MgF\textsubscript{2} and UVFS windows is shown in Fig.\,\ref{fig:UVFS}, compared to the manufacturer's transmission curves. 
There is a good match in the position of the filter edges and in transmittance values at longer wavelengths, yet some deviations are also visible at shorter wavelengths. 
This difference potentially originates from surface contamination and scattering caused by particles that are partially visible in the insets as circular rings.

\section{\label{sec:CCD} CCD camera calibration}
To calibrate the CCD quantum efficiency, we used the same principle as for the photodiode in section \ref{sec:PD}, i.e. the CCD and the absolutely calibrated beamline photodiode are exposed by the same photon flux. The photon flux $\phi$ can be measured in units of \unit{photons/s} by the beamline photodiode as: 
\begin{equation} \label{eq:phi}
     \phi=\frac{I_{ref}}{R_{ref}\cdot{q}\cdot{E_{ph}}}
\end{equation}
where $I_{ref}$ and $R_{ref}$ represent, respectively, the current and spectral responsivity of the reference photodiode,  while $q$ represents the elementary charge to convert the energy unit from J to eV, and $E_{ph}$ corresponds to the photon energy in eV units.

When such a stream of photons is exposed to a CCD camera, it goes through three main conversion stages. 
In the first stage, a fraction of the incident photons is detected, represented by the quantum efficiency (QE). 
In the second step, the detected photons are converted to electrons according to the quantum yield (QY) of the CCD. 
For photons with an energy higher than the indirect band-gap of Si, a photon energy of 3.66\,eV is needed to generate an electron\cite{scholze2000, heymes20202}. 
This implies that more than one electron can be generated per detected photon for photon energies higher than 3.66\,eV, corresponding to wavelengths below 339\,nm. 
In this range, the number of electrons generated can be calculated as $E_{ph} / 3.66$.
For wavelengths longer than 339\,nm this assumption no longer holds, and QY converges to a constant value of 1. Hence, QY equals $1/3.66$\,\textit{e}\textsuperscript{-}/eV for wavelengths shorter than 339\,nm and it is unity for longer wavelengths~\cite{Kuschnerus1998}. 
In the last stage, electrons are converted to a CCD signal in the arithmetic digital unit (ADU) according to the gain $G$ of the CCD camera. The gain has a unit of \textit{e}\textsuperscript{-}/ADU. 
The signal $S$ generated by a CCD camera that is exposed to the same photon flux as the photodiode can be written as: 
\begin{equation} \label{eq:ccd}
S = \phi \cdot \frac{ QE \cdot QY}{G}\cdot t
\end{equation}
where $t$ is the exposure duration of the CCD camera in units of seconds.
Thus, quantum efficiency can be derived from Eqs.\,\ref{eq:phi} and \ref{eq:ccd} as: 
\begin{equation} \label{eq:qe}
    QE = \frac{S \cdot G}{t \cdot QY} \cdot
    \frac{R_{ref} \cdot q \cdot E_{ph}}{I_{ref}}.
\end{equation}
Again, the time-dependent photon flux variations are taken into account by normalizing both photodiode and CCD signals to the monitor currents, but are not presented here. 

\begin{figure}
\includegraphics[width=\columnwidth]{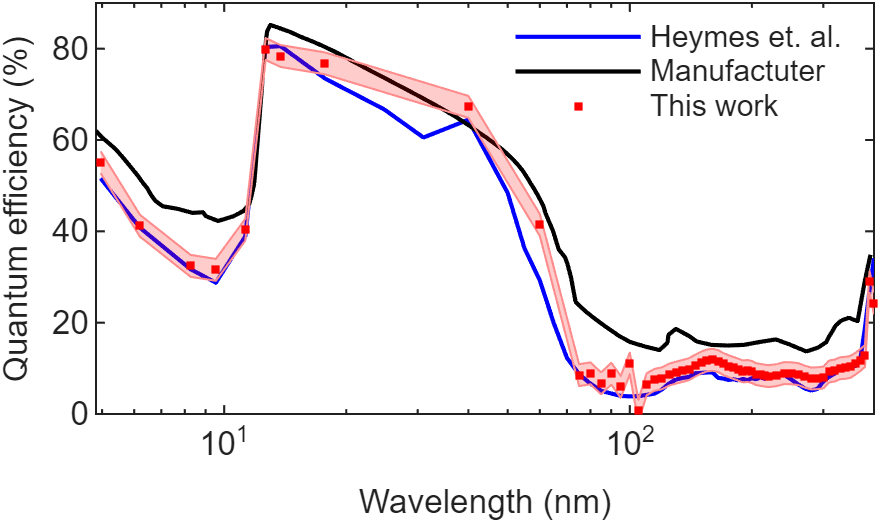}%
\caption{\label{fig:qe} 
Measured quantum efficiency of the CCD camera, compared to Heymes~et.al.\cite{heymes20201} and manufacturer\cite{princeton2016}.}
\end{figure}

The measured quantum efficiency of the CCD is shown in~Fig.\,\ref{fig:qe} compared to two representative curves from the literature. First of all, there is a large difference between the literature reports\cite{princeton2016, heymes20201} almost in all wavelength ranges. 
Our measurement is mainly consistent with the report of Heymes~et.al.\cite{heymes20201} that is measured experimentally in a synchrotron. Less information is available on the measurement conditions of the manufacturer QE\cite{princeton2016}. 
The most noticeable difference appears in the EUV range below 12\,nm and in the VUV/DUV wavelengths longer than 50\,nm, both experimentally measured QE curves being lower than the manufacturer QE. 
The difference may arise from different processing conditions and the contamination of the specific CCD chips used. 
The photon absorption depth in the 6--12\,nm and 100-300\,nm ranges is known to be less than 100\,nm and 10\,nm, respectively. 
In the 12--100\,nm range the absorption depth ranges between 300-1000\,nm\cite{moody2017, heymes20201}. 
In spectral ranges with shallow absorption depths less than 100\,nm, the QE can decrease as a result of additional absorption in native silicon oxide and typical carbon contamination layers on CCD chips.

\section{\label{sec:gratings} Diffraction efficiency of gratings}
For calibration of the gratings, the CCD of the spectrometer has been used as the detector to eliminate uncertainties that may originate from mounting of the gratings and other optical components. 
For each grating, two datasets have been recorded: one with the grating on the beam path and one without it to serve as the reference signal. 
The same spectrometer slit has been used in both datasets to define the same solid angle for the incoming beam. 
The camera images when the grating is on the beam path include several diffraction orders that are separated spatially. 
The distance between them depends on the wavelength of the photon beam and follows the grating equation. 
The diffraction orders have been individually identified by image processing, and the total signals for each order have been calculated. 
The efficiency of the diffraction order $m$ has been calculated as 
\begin{equation}
    \eta_m=\frac{S_m - S_b}{S_{ref} - S_b}
\end{equation} 
where $S_m$ is the diffraction order signal, $S_{ref}$ is the reference signal, $S_b$ is the background signal and $m=0,1,2,3…$ is the diffraction order number. 
This work presents calibration of gratings with 3 different line densities, namely \num{10000}\,lines/mm (for EUV range), \num{3000}\, lines/mm (for VUV range) and \num{1000}\,lines/mm (for DUV range) for horizontal \textit{s-}polarization, unless otherwise stated. 
The gratings can be considered as amplitude gratings in the studied range due to the high absorption in the chosen material stack. 
More detailed specifications of the gratings and their fabrication process can be found in\cite{goh2015, bayraktar2016}. 

\subsubsection{\label{sec:10k} EUV grating}
\begin{figure}
\includegraphics[width=\columnwidth]{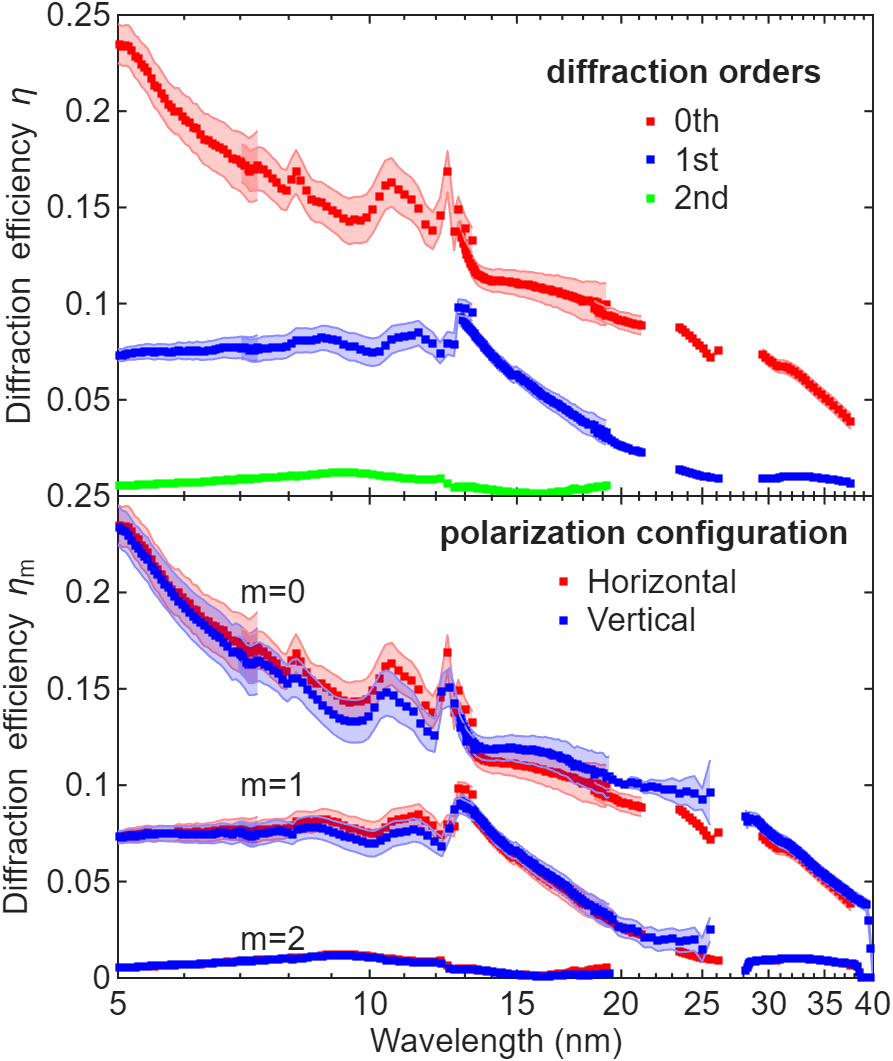}%
\caption{\label{fig:10k} Diffraction efficiency of \num{10000}\,lines/mm grating for horizontal (\textit{s}) polarization (top). Diffraction efficiency for both polarizations (bottom).}
\end{figure}

The grating with \num{10000}\,lines/mm is designed for measurements in the EUV range up to 40\,nm with the highest spectral resolution. 
This grating is typically used with a 50\,\textmu m wide slit in front of it, so it is calibrated with this slit. 
The diffraction efficiencies for the zeroth, first, and second diffraction orders are presented in Fig.\,\ref{fig:10k}. 
The error bands in this and other grating efficiency figures are calculated using error propagation assuming an uncertainty of 3\% of the maximum of the signals $S_m$ and $S_{ref}$ in each filter range.  
The overlap of some datapoints, such as around 7\,nm can be observed due to the beamline's filter edges. 
In combining the efficiency data in different filter ranges, some mismatches in the overlapping energies have been observed. 
Such a mismatch is observed least for the shorter wavelengths, hence data of filters at longer wavelengths were scaled with a constant value to match the data of the filter at the shortest wavelength. 

The obtained diffraction efficiency curve for the first order, which is relevant for spectroscopy purposes, shows a flat response in the 7-8\% range up to 12\,nm, around which a sharp increase is observed. This sharp feature can be ascribed to the Si-L absorption edge due to the SiN layer that is used in the grating stack\cite{goh2015, bayraktar2016}. 
Beyond this wavelength, the efficiency decreases. 
The zero order shows an overall decreasing trend with wavelength, whereas the second order is flat around 1\%. 
In Fig.~\ref{fig:10k} we also compare the diffraction efficiencies for different polarizations. 
There is a good match between the efficiency curves for two polarizations, indicating a negligible dependence on polarization for the grating. 

\subsubsection{\label{sec:1,3k} VUV and DUV gratings}
\begin{figure}
\includegraphics[width=\columnwidth]{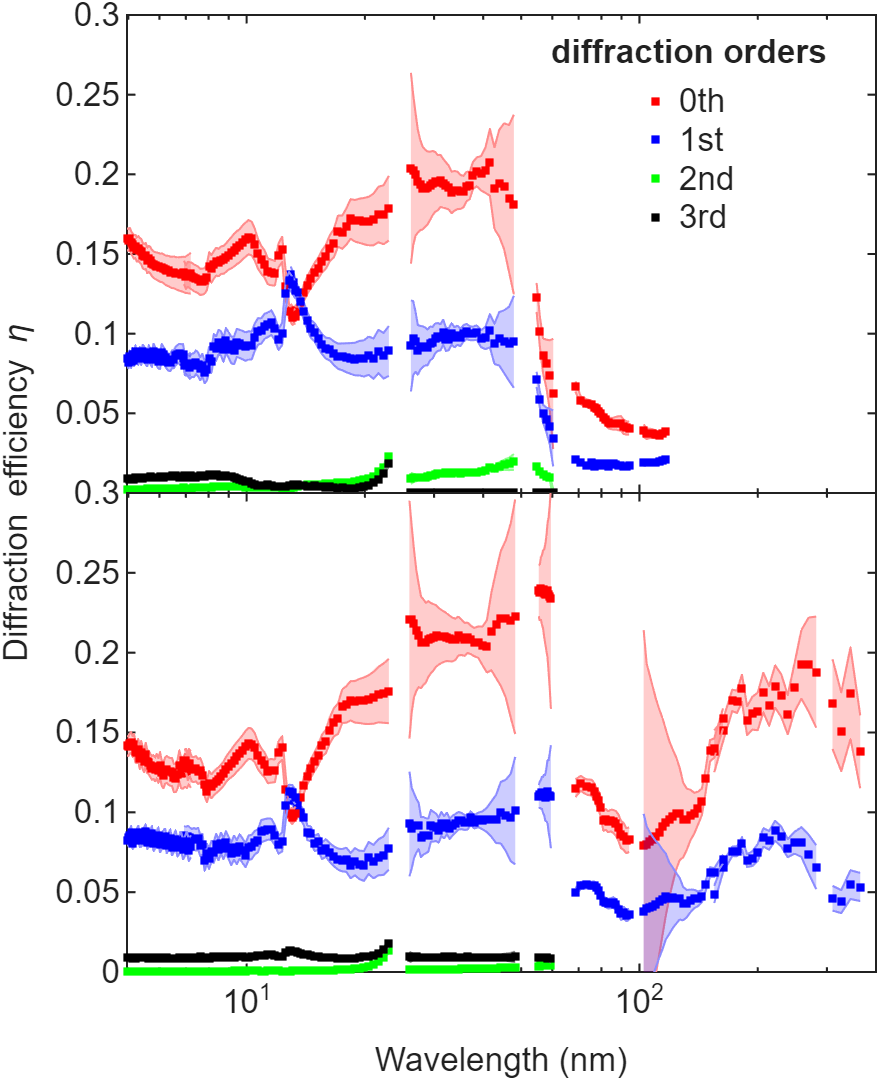}%
\caption{\label{fig:3k}Diffraction efficiency of \num{3000}\,lines/mm (top), \num{1000}\,lines/mm (bottom) gratings.}
\end{figure}

The grating with \num{3000}\,lines/mm is designed for VUV measurements for wavelengths up to 120\,nm and \num{1000}\,lines/mm is designed for DUV measurements up to 400\,nm. We present diffraction efficiencies in Fig.~\ref{fig:3k} that are obtained using a 150\,\textmu m slit. 
For the first order, the flat efficiency curve is again visible up to 50 and 60\,nm for VUV and DUV gratings, respectively. 
Like the EUV grating, the efficiency fluctuates around 8\% for the VUV and DUV gratings. 
The sharp feature around 12\,nm is again visible for both gratings, as expected due to Si-L absorption edge. 
For longer wavelengths, the efficiency first decreases rapidly for both gratings and increases again for the DUV grating after about 150\,nm. 
The zero order efficiency shows a variation in the 5--20\% range for both gratings. 
The second and third order efficiencies are below 1\% level for both gratings. 
In general, uncertainty in diffraction efficiency increases due to a lower signal-to-noise ratio in some ranges, especially in the 25-50\,nm range and around 100\,nm.

\subsubsection{Discussion}
The overall trend of the efficiencies can be explained by considering the ratio of  wavelength to the period of the gratings. 
For square-wave amplitude gratings with period $d$ much larger than the wavelength $\lambda$ ($d \gg \lambda$), the diffraction efficiency according to scalar diffraction theory is expected to be independent of the wavelength and can be written as\cite{harvey2019}:
\begin{equation}
    \eta_m = \left(\frac{b}{d}\right)^2\mathrm{sinc}^2{\left(\frac{m b}{d}\right)}
\end{equation}
where $b$ is the width of the open slits between the grating bars. 
As observed in all grating efficiencies, flat efficiency curves tend to reduce at longer wavelengths, where the validity of the $d \gg \lambda$ condition starts to diminish. 
The validity range is widest for the grating with the largest period, i.e. the lowest density grating (\num{1000}\,lines/mm). This grating has a flat efficiency curve that goes up to 60\,nm which is the longest wavelength in all gratings. 

For $b/d = 0.5$, the efficiencies of the zeroth, first, second, and third diffraction orders are expected to be 25\%, 10\%, 0 and 1\% respectively\cite{harvey2019}. 
Considering that 20\% of the area of the gratings are covered by the support bars, the efficiencies are expected to be lowered by 20\% in a first order approximation. 
This consideration explains the observed 8\% first order efficiency of the gratings at short wavelengths. 

Any deviation from $b/d = 0.5$ results in a decrease in the efficiency of the first order and an increase in the efficiency of the second order. 
The EUV grating has a slightly lower first order efficiency compared to VUV and DUV gratings at short wavelengths. 
This reduction can come from a slight deviation from 0.5 in the $b/d$ ratio due to fabrication tolerance becoming non-negligible relative to the fine 50\,nm bar width in the \num{10000}\,lines/mm grating.  
Second order efficiency is also highest in the EUV grating, which also supports this hypothesis. 

\section{\label{sec:Efficiency} Spectrometer efficiency}
\begin{figure}[tb]
\includegraphics[width=\columnwidth]{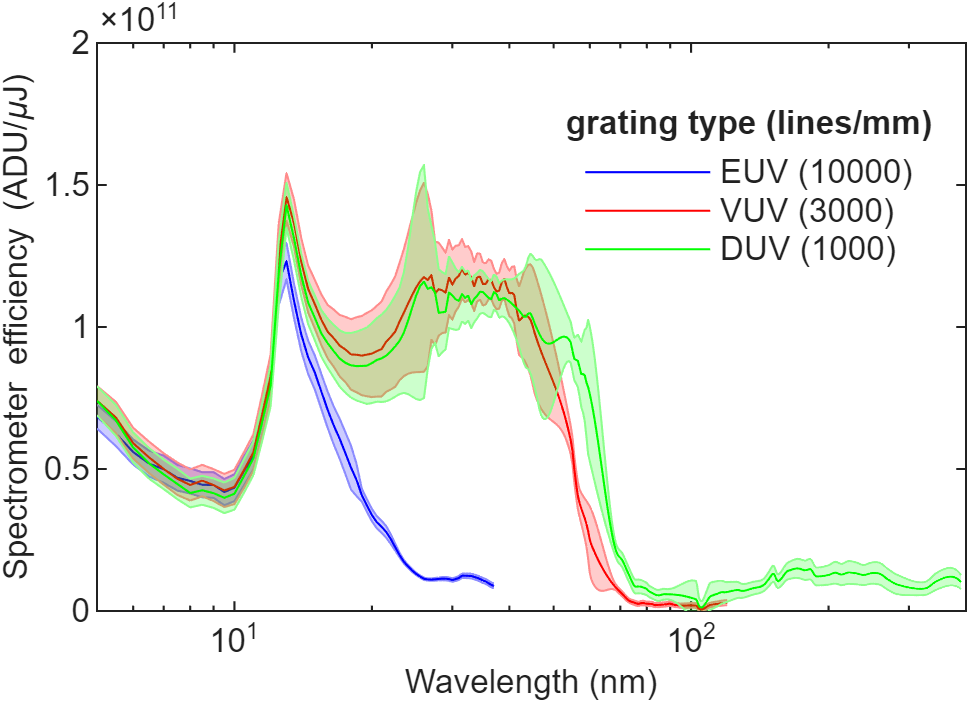}%
\caption{\label{fig:eff}Spectrometer efficiency using three grating configurations and assuming a unity camera gain.}
\end{figure}

For spectral measurements, the overall efficiency of the spectrometer as a function of wavelength is the most relevant parameter. 
Several combinations of spectrometer components such as filters, gratings, and parameters such as camera gain are possible; however, a grating coupled with a CCD camera is required at minimum. 
Here we report the spectrometer efficiency based on the first order diffraction efficiencies of three gratings as shown in Fig.~\ref{fig:eff}, assuming unity camera gain. 
The efficiencies are interpolated using Modified Akima interpolation. 
It can be seen that there is an overall matching of the efficiencies of three gratings at short wavelengths, but for wavelengths longer than 12\,nm the VUV and DUV gratings differ significantly from the EUV grating . 
The difference can be explained by the larger period of the VUV and DUV gratings compared to the EUV grating, considering that a flat efficiency curve independent of grating period is expected for a square-wave amplitude grating but the validity of this prediction diminishes as the wavelength to period ratio increases.  

\section{\label{sec:Conclusion} Conclusion}

In this work, we present the absolute, component-by-component calibration of a TGS with uncertainty estimates. 
The spectrometer components, including filters, gratings, a photodiode, and a CCD camera, were calibrated using monochromatic synchrotron radiation as the light source and an absolutely calibrated photodiode as the reference standard. 
Absolute calibration for various combinations of spectrometer components can be derived from the data presented, providing considerable versatility. 
This broadband calibration in the 5-400\,nm range can be useful for various spectroscopy studies in areas such as EUV lithography, solar corona studies, plasma diagnostics and spectral metrology of optical components. 

\begin{acknowledgments}
This work was partly financed by ASML B.V., Horizon Europe, and Rijksdienst voor Ondernemend (RVO) Nederland under the project 14 Anstrom CMOS IC technology (Grant agreement ID: 101096772). 
We acknowledge Elettra Sincrotrone Trieste for providing access to its synchrotron radiation facility.
\end{acknowledgments}

\section*{Data Availability Statement}
Data are available upon reasonable request. 

\section{\label{sec:refs} References}

\bibliography{refs} 

@PREAMBLE{
 "\providecommand{\noopsort}[1]{}" 
 # "\providecommand{\singleletter}[1]{#1}%" 
}

@techreport{moody2017,
    author = {Moody, Ian and Watkins, Marc and Bell, Ray and Soman, Matthew and Keelan, Jonathan and Holland, Andrew},
    title = {{CCD QE} in the Soft X-ray Range},
    institution = {e2v Technologies},
    year = 2017
}

@manual{princeton2016,
    organization = "Princeton Instruments",
    title = "{PIXIS-XO: 2KB} Datasheet",
    number = "{PIXIS-XO:2KB} Rev. P1",
    year =  2016,
    month =  6,    
    note = "{PIXIS-XO:2KB} Rev. P1"
}

@article{ye2026,
    author = {Ye, Z. and Zhang, Z. and Lan, T. and Ma, H. and Wu, J. and Sokolov, A. and Yu, Y. and Wang, Z.},
    title = {A transmission grating extreme ultraviolet spectrometer using a gradient-reflectivity Al/Zr multilayer mirror for the range of 24–32 nm},
    journal = {Review of Scientific Instruments},
    volume = {97},
    number = {5},
    pages = {053506},
    year = {2026},
    month = {05},
    issn = {0034-6748},
    doi = {10.1063/5.0317163},
    url = {https://doi.org/10.1063/5.0317163},
}

@article{Feng1997ATG,
  title={A transmission grating spectrograph and its application in both laser plasma and plasma focus},
  author={X. Feng and Mahe Liu and Sing Lee},
  journal={Review of Scientific Instruments},
  year={1997},
  volume={68},
  pages={3068-3073},
  url={https://api.semanticscholar.org/CorpusID:120280017}
}

@article{Fujikawa1998InSC,
  title={In situ calibration of a flat-field grazing incidence spectrograph with an x-ray charge-coupled device camera using a transmission grating and a laser-produced plasma for x-ray laser research},
  author={Chiemi Fujikawa and Naohiro Yamaguchi and Tamio T. Hara and Tetsuya Kawachi and Hitoshi Oyama and Kōzō Andō and Yoshinobu Aoyagi},
  journal={Review of Scientific Instruments},
  year={1998},
  volume={69},
  pages={2849-2853},
  url={https://api.semanticscholar.org/CorpusID:120918539}
}

@article{Sailaja1998ASX,
  title={A simple XUV transmission grating spectrograph with sub-{\aa}ngstr{\"o}m resolution for laser-plasma interaction studies},
  author={S Sailaja and V Arora and S R Kumbhare and P A Naik and P D Gupta and D A Fedin and A A Rupasov and A S Shikanov},
  journal={Measurement Science and Technology},
  year={1998},
  volume={9},
  pages={1462 - 1468},
  url={https://api.semanticscholar.org/CorpusID:250899985}
}

@article{Wilhein1999ASG,
  title={A slit grating spectrograph for quantitative soft x-ray spectroscopy},
  author={Thomas Wilhein and Stefan Rehbein and Dirk Hambach and Magnus Berglund and Lars Rymell and Hans M. Hertz},
  journal={Review of Scientific Instruments},
  year={1999},
  volume={70},
  pages={1694-1699},
  url={https://api.semanticscholar.org/CorpusID:16235071}
}

@article{Weaver2001TheDO,
  title={The determination of absolutely calibrated spectra from laser produced plasmas using a transmission grating spectrometer at the Nike laser facility},
  author={J. l. Weaver and Glenn Holland and Uri Feldman and John F Seely and Catherine Brown and Victor Serlin and Alban Veli Deniz and Marcel Klapisch},
  journal={Review of Scientific Instruments},
  year={2001},
  volume={72},
  pages={108-118},
  url={https://api.semanticscholar.org/CorpusID:121236645}
}

@article{Seely2001CalibrationOA,
  title={Calibration of an extreme-ultraviolet transmission grating spectrometer with synchrotron radiation.},
  author={John F Seely and Charles M. Brown and Glenn Holland and Frederick A. Hanser and J. Wise and J. l. Weaver and Raj S. Korde and Rodney A. Viereck and Richard Grubb and Darrell L. Judge},
  journal={Applied optics},
  year={2001},
  volume={40},
  number={10},
  pages={1623-30},
  url={https://api.semanticscholar.org/CorpusID:17370283}
}

@article{Hatsui2005DesignOA,
  title={Design of a novel transmission-grating spectrometer for soft X-ray emission studies},
  author={Takaki Hatsui and Hiroyuki Setoyama and Eiji Shigemasa and Nobuhiro Kosugi},
  journal={Journal of Electron Spectroscopy and Related Phenomena},
  year={2005},
  volume={144},
  pages={1059-1062},
  url={https://api.semanticscholar.org/CorpusID:121940523}
}

@article{Kornilov2010NanogratingbasedCV,
  title={Nanograting-based compact vacuum ultraviolet spectrometer and beam profiler for in situ characterization of high-order harmonic generation light sources.},
  author={Oleg Kornilov and Russell B. Wilcox and Oliver Gessner},
  journal={The Review of scientific instruments},
  year={2010},
  volume={81 6},
  pages={063109},
  url={https://api.semanticscholar.org/CorpusID:34620693}
}

@article{banine2011,
  author = {Banine, V. Y. and Koshelev, K. N. and Swinkels, G. H. P. M.},
  title = {{Physical processes in EUV sources for microlithography}},
  journal = {J. Phys. D},
  volume = {44},
  number = {25},
  pages = {253001},
  year = {2011}
}

@manual{AXUV_datasheet,
  title        = {Photodiode D5 mm AXUV20HS1 Datasheet},
  organization = {Opto Diode Corp},
  year         = {2019},
  url          = {https://www.digikey.com/en/products/detail/opto-diode-corp/AXUV20HS1/4486136},
}

@phdthesis{thesisTryus,
  author    = {Tryus, Maksym},
  title     = {Extreme ultraviolet reflectometry for structural and optical characterization of thin films and layer systems},
  school    = {RWTH Aachen University, University of Padova
},
  year      = {2018},
  type      = {PhD thesis},
  address   = {Aachen, Germany}
}

@article{bayraktar2016,
  author = {Bayraktar, Muharrem and Bastiaens, Hubertus M J and Bruineman, Caspar and Vratzov, Boris and Bijkerk, Frederik},
  title = {{Broadband transmission grating spectrometer for measuring the emission spectrum of EUV sources}},
  journal = {NEVAC blad},
  volume = {54},
  number = {1},
  pages = {14--19},
  year = {2016},
  publisher = {Nederlandse Vacuumvereniging},
  issn = {0169-9431},
  language = {English}
}

@book{born1999,
  author = {Born, Max and Wolf, Emil},
  title = {Principles of Optics: Electromagnetic Theory of Propagation, Interference and Diffraction of Light},
  year = {1999},
  publisher = {Cambridge University Press},
  isbn = {9780521642224},
  edition = {7th}
}

@article{bouza2020,
  author = {Z Bouza and J Scheers and A Ryabtsev and R Schupp and L Behnke and C Shah and J Sheil and M Bayraktar and J R Crespo L{\'{o}}pez-Urrutia and W Ubachs and R Hoekstra and O O Versolato},
  title = {{EUV} spectroscopy of {Sn}\textsuperscript{5+}-{Sn}\textsuperscript{10+}  ions in an electron beam ion trap and laser-produced plasmas},
  journal = {J. Phys. B: At. Mol. Opt. Phys.},
  volume = {53},
  number = {19},
  pages = {195001},
  year = {2020},
  month = {aug},
  publisher = {{IOP} Publishing}
}

@article{bouza2021,
  author = {Bouza, Z and Byers, J and Scheers, J and Schupp, R and Mostafa, Y and Behnke, L and Mazzotta, Z and Sheil, J and Ubachs, W and Hoekstra, R and others},
  title = {{The spectrum of a 1-$\mu$ m-wavelength-driven tin microdroplet laser-produced plasma source in the 5.5--265.5 nm wavelength range}},
  journal = {AIP Adv.},
  volume = {11},
  number = {12},
  pages = {125003},
  year = {2021},
  month = {12},
  publisher = {AIP Publishing LLC},
  doi = {10.1063/5.0073839},
  issn = {2158-3226},
  url = {https://doi.org/10.1063/5.0073839}
}

@article{burcklen2024,
  author = {Burcklen, Catherine and Delmotte, Franck and Alameda, Jennifer and Salmassi, Farhad and Gullikson, Eric and Soufli, Regina},
  title = {Optical constants of magnetron sputtered aluminum in the range 17–1300 eV with improved accuracy and ultrahigh resolution in the L absorption edge region},
  journal = {Journal of Applied Physics},
  volume = {136},
  number = {19},
  pages = {195106},
  year = {2024},
  month = {11},
  doi = {10.1063/5.0233781},
  issn = {0021-8979},
  url = {https://doi.org/10.1063/5.0233781}
}

@article{churilov20061,
  author = {Churilov, S S and Ryabtsev, A N},
  title = {{Analysis of the $4p^{6}$ $4d^{7}$--($4p^{6}$ $4d^{6}4f$ + $4p^{5}$ $4d^{8}$) transitions in the Sn VIII spectrum}},
  journal = {Opt. Spectrosc.},
  volume = {100},
  number = {5},
  pages = {660--666},
  year = {2006},
  publisher = {Springer}
}

@article{churilov20062,
  author = {Churilov, S S and Ryabtsev, A N},
  title = {{Analyses of the Sn IX--Sn XII spectra in the EUV region}},
  journal = {Phys. Scr.},
  volume = {73},
  number = {6},
  pages = {614},
  year = {2006},
  publisher = {IOP Publishing}
}

@article{churilov20063,
  author = {Churilov, S S and Ryabtsev, A N},
  title = {{Analysis of the Spectra of In XII--XIV and Sn XIII--XV in the Far--VUV Region}},
  journal = {Opt. Spectr.},
  volume = {101},
  number = {2},
  pages = {169--178},
  year = {2006},
  publisher = {Springer}
}

@article{gambino2015,
  author = {Nadia Gambino and Bob Rollinger and Duane Hudgins and Reza   S. Abhari},
  title = {{Spectral emission properties of a laser-produced plasma light source in the sub-200 nm range for wafer inspection applications}},
  journal = {J. Micro/Nanolithogr. MEMS MOEMS},
  volume = {14},
  number = {3},
  pages = {1 -- 7},
  year = {2015},
  publisher = {SPIE}
}

@article{gambino2016,
  author = {Nadia Gambino and Bob Rollinger and Markus Brandstätter and Reza S. Abhari},
  title = {{Vacuum ultraviolet spectral emission properties of {Ga}, {In} and {Sn} droplet-based laser produced plasmas}},
  journal = {Spectrochim. Acta B: At. Spectrosc.},
  volume = {122},
  pages = {149--156},
  year = {2016},
  issn = {0584-8547}
}

@article{goh2015,
  author = {S.J. Goh and H.J.M. Bastiaens and B Vratzov and Q. Huang and F. Bijkerk and K.J. Boller},
  title = {Fabrication and characterization of free-standing, high-line-density transmission gratings for the vacuum UV to soft X-ray range},
  journal = {Opt. Express},
  volume = {23},
  number = {4},
  pages = {4421--4434},
  year = {2015},
  month = {Feb},
  publisher = {Optica Publishing Group},
  doi = {10.1364/OE.23.004421},
  url = {https://opg.optica.org/oe/abstract.cfm?URI=oe-23-4-4421}
}

@article{harvey2019,
  author = {James E. Harvey and Richard N. Pfisterer},
  title = {{Understanding diffraction grating behavior: including conical diffraction and Rayleigh anomalies from transmission gratings}},
  journal = {Optical Engineering},
  volume = {58},
  number = {8},
  pages = {087105},
  year = {2019},
  publisher = {SPIE},
  doi = {10.1117/1.OE.58.8.087105},
  url = {https://doi.org/10.1117/1.OE.58.8.087105}
}

@article{henke1993,
  author = {B.L. Henke and E.M. Gullikson and J.C. Davis},
  title = {{X-Ray Interactions: Photoabsorption, Scattering, Transmission, and Reflection at {E} = 50-30,000 eV, {Z} = 1-92}},
  journal = {At. Data Nucl. Data Tables},
  volume = {54},
  number = {2},
  pages = {181--342},
  year = {1993},
  issn = {0092-640X}
}

@article{heymes20201,
  author = {J. {Heymes} and M. {Soman} and G. {Randall} and A. {Gottwald} and A. {Harris} and A. {Kelt} and I. {Moody} and X. {Meng} and A. D. {Holland}},
  title = {Comparison of Back-Thinned Detector Ultraviolet Quantum Efficiency for Two Commercially Available Passivation Treatments},
  journal = {IEEE Trans. Nucl. Sci.},
  volume = {67},
  number = {8},
  pages = {1962--1967},
  year = {2020}
}

@inproceedings{heymes20202,
  author = {Julian Heymes and Matthew Soman and Thomas Buggey and Chiaki Crews and George Randall and Alexander Gottwald and Andrew Harris and Andrew Kelt and Udo Kroth and Ian Moody and Xiao Meng and Oskar Ogor and Andrew Holland},
  title = {{Calibrating Teledyne-e2v’s ultraviolet image sensor quantum efficiency processes}},
  booktitle = {X-Ray, Optical, and Infrared Detectors for Astronomy IX},
  volume = {11454},
  pages = {263 -- 273},
  year = {2020},
  publisher = {SPIE},
  editor = {Andrew D. Holland and James Beletic},
  organization = {International Society for Optics and Photonics}
}

@article{horst2016,
  author = {R M van der Horst and J Beckers and E A Osorio and D I Astakhov and W J Goedheer and C J Lee and V V Ivanov and V M Krivtsum and K N Koshelev and D V Lopaev and F Bijkerk and V Y Banine},
  title = {{Exploring the electron density in plasma induced by {EUV} radiation: I. Experimental study in hydrogen}},
  journal = {J. Phys. D: Appl. Phys.},
  volume = {49},
  number = {14},
  pages = {145203},
  year = {2016},
  month = {mar},
  publisher = {{IOP} Publishing}
}

@article{huang2017,
  author = {Qiushi Huang and Viacheslav Medvedev and Robbert van de Kruijs and Andrey Yakshin and Eric Louis and Fred Bijkerk},
  title = {{Spectral tailoring of nanoscale {EUV} and soft x-ray multilayer optics}},
  journal = {Appl. Phys. Rev.},
  volume = {4},
  number = {1},
  pages = {011104},
  year = {2017},
  month = {03},
  doi = {10.1063/1.4978290},
  issn = {1931-9401}
}

@inproceedings{joseph20131,
  author = {Aby Joseph and Mohammed H. Modi and Amol Singh and R. K. Gupta and G. S. Lodha},
  title = {{Analysis of soft x-ray/VUV transmission characteristics of Si and Al filters}},
  booktitle = {AIP Conference Proceedings},
  volume = {1512},
  pages = {498--499},
  year = {2013},
  publisher = {American Institute of Physics}
}

@article{kerkhof20201,
  author = {Mark A. van de Kerkhof and Fei Liu and Marieke  Meeuwissen and Xueqing Zhang and Muharrem Bayraktar and Robert C. de Kruif and Natalia V. Davydova},
  title = {{High-power EUV lithography: spectral purity and imaging performance}},
  journal = {J. Micro/Nanolithogr. MEMS MOEMS},
  volume = {19},
  number = {3},
  pages = {1 -- 16},
  year = {2020},
  publisher = {SPIE}
}

@inproceedings{kerkhof20202,
  author = {Mark van de Kerkhof and Fei Liu and Marieke Meeuwissen and Xueqing Zhang and Robert de Kruif and Natalia Davydova and Guido Schiffelers and Felix Wählisch and Eelco van Setten and Wouter Varenkamp and Kees Ricken and Laurens de Winter and John McNamara and Muharrem Bayraktar},
  title = {{Spectral purity performance of high-power EUV systems}},
  booktitle = {Extreme Ultraviolet (EUV) Lithography XI},
  volume = {11323},
  pages = {435 -- 450},
  year = {2020},
  publisher = {SPIE},
  editor = {Nelson M. Felix and Anna Lio},
  organization = {International Society for Optics and Photonics}
}

@article{louis2011,
  author = {E. Louis and A.E. Yakshin and T. Tsarfati and F. Bijkerk},
  title = {{Nanometer interface and materials control for multilayer EUV-optical applications}},
  journal = {Progress in Surface Science},
  volume = {86},
  number = {11},
  pages = {255--294},
  year = {2011},
  issn = {0079-6816}
}

@book{mackay2020,
  author = {Tom G. Mackay and Akhlesh Lakhtakia},
  title = {The Transfer-Matrix Method in Electromagnetics and Optics},
  year = {2020},
  publisher = {Springer Nature}
}

@article{nannarone2004,
  author = {Nannarone, S. and Borgatti, F. and DeLuisa, A. and Doyle, B. P. and Gazzadi, G. C. and Giglia, A. and Finetti, P. and Mahne, N. and Pasquali, L. and Pedio, M. and Selvaggi, G. and Naletto, G. and Pelizzo, M. G. and Tondello, G.},
  title = {The BEAR Beamline at Elettra},
  journal = {AIP Conference Proceedings},
  volume = {705},
  pages = {450--453},
  year = {2004},
  doi = {10.1063/1.1757831}
}

@article{osullivan2015,
  author = {O'Sullivan, Gerry and Li, Bowen and D'Arcy, Rebekah and Dunne, Padraig and Hayden, Paddy and Kilbane, Deirdre and McCormack, Tom and Ohashi, Hayato and O’Reilly, Fergal and Sheridan, Paul and others},
  title = {{Spectroscopy of highly charged ions and its relevance to EUV and soft x-ray source development}},
  journal = {J. Phys. B},
  volume = {48},
  number = {14},
  pages = {144025},
  year = {2015},
  month = {may},
  publisher = {IOP Publishing},
  doi = {10.1088/0953-4075/48/14/144025},
  url = {https://doi.org/10.1088/0953-4075/48/14/144025}
}

@book{palik1985,
  author = {Palik, Edward D},
  title = {Handbook of optical constants of solids},
  volume = {1},
  year = {1985},
  publisher = {Academic press}
}

@article{pasquali2004,
  author = {L. Pasquali and A. De Luisa and S. Nannarone},
  title = {The UHV Experimental Chamber for Optical Measurements and Angle Resolved Photoemission of the BEAR Beamline at ELETTRA},
  journal = {AIP Conf. Proc.},
  volume = {705},
  pages = {1142--1145},
  year = {2004},
  doi = {10.1063/1.1758001}
}

@article{powell1990,
  author = {Forbes R. Powell and Peter W. Vedder and Joakim F. Lindblom and Stephen F. Powell},
  title = {{Thin film filter performance for extreme ultraviolet and x-ray applications}},
  journal = {Opt. Eng.},
  volume = {29},
  number = {6},
  pages = {614 -- 624},
  year = {1990},
  publisher = {SPIE}
}

@article{scholze2000,
  author = {F. Scholze and H. Henneken and P. Kuschnerus and H. Rabus and M. Richter and G. Ulm},
  title = {Determination of the electron–hole pair creation energy for semiconductors from the spectral responsivity of photodiodes},
  journal = {Nuclear Instruments and Methods in Physics Research Section A: Accelerators, Spectrometers, Detectors and Associated Equipment},
  volume = {439},
  number = {2},
  pages = {208--215},
  year = {2000},
  doi = {https://doi.org/10.1016/S0168-9002(99)00937-7},
  issn = {0168-9002},
  url = {https://www.sciencedirect.com/science/article/pii/S0168900299009377}
}

@article{torretti2017,
  author = {Torretti, F. and Windberger, A. and Ryabtsev, A. and Dobrodey, S. and Bekker, H. and Ubachs, W. and Hoekstra, R. and Kahl, E. V. and Berengut, J. C. and Crespo L\'opez-Urrutia, J. R. and Versolato, O. O.},
  title = {{Optical spectroscopy of complex open-$4d$-shell ions ${\mathrm{Sn}}^{7+}-{\mathrm{Sn}}^{10+}$}},
  journal = {Phys. Rev. A},
  volume = {95},
  pages = {042503},
  year = {2017},
  month = {Apr},
  publisher = {American Physical Society},
  issue = {4},
  numpages = {16}
}

@article{torretti2018,
  author = {Torretti, F and Schupp, Ruben and Kurilovich, Dmitry and Bayerle, Alex and Scheers, Joris and Ubachs, Wim and Hoekstra, Ronnie and Versolato, O O},
  title = {{Short-wavelength out-of-band EUV emission from Sn laser-produced plasma}},
  journal = {J. Phys. B},
  volume = {51},
  number = {4},
  pages = {045005},
  year = {2018},
  publisher = {IOP Publishing}
}

@article{torretti2019,
  author = {Torretti, Francesco and Liu, Fei and Bayraktar, Muharrem and Scheers, Joris and Bouza, Zoi and Ubachs, W and Hoekstra, R and Versolato, Oscar},
  title = {{Spectral characterization of an industrial EUV light source for nanolithography}},
  journal = {J. Phys. D},
  volume = {53},
  number = {5},
  pages = {055204},
  year = {2019},
  publisher = {IOP Publishing}
}

@article{torretti2020,
  author = {Torretti, F. and Sheil, J. and Schupp, R. and Basko, M. M. and Bayraktar, M. and Meijer, R. A. and Witte, S. and Ubachs, W. and Hoekstra, R. and Versolato, O. O. and Neukirch, A. J. and Colgan, J.},
  title = {{Prominent radiative contributions from multiply-excited states in laser-produced tin plasma for nanolithography}},
  journal = {Nat. Commun.},
  volume = {11},
  number = {1},
  pages = {1--8},
  year = {2020},
  month = may,
  day = {11},
  publisher = {Nature Publishing Group},
  doi = {10.1038/s41467-020-15678-y},
  issn = {2041-1723},
  language = {English}
}

@article{versolato2019,
  author = {Versolato, Oscar O},
  title = {{Physics of laser-driven tin plasma sources of {EUV} radiation for nanolithography}},
  journal = {Plasma Sourc. Sci. Tech.},
  volume = {28},
  number = {8},
  pages = {083001},
  year = {2019},
  month = {sep},
  publisher = {{IOP} Publishing},
  doi = {10.1088/1361-6595/ab3302},
  url = {https://doi.org/10.1088/1361-6595/ab3302}
}

@article{Kuschnerus1998,
  author  = {P. Kuschnerus and H. Rabus and M. Richter and F. Scholze and L. Werner and G. Ulm},
  title   = {Characterization of photodiodes as transfer detector standards in the 120 nm to 600 nm spectral range},
  journal = {Metrologia},
  year    = {1998},
  volume  = {35},
  number  = {4},
  pages   = {355},
  doi     = {10.1088/0026-1394/35/4/23}
}

\end{document}